\documentclass[useAMS,usenatbib]{mn2e}
\usepackage[utf8]{inputenc}
\usepackage{graphicx}
\title[Accretion disks spectra and ULXs]{A comparative analysis of standard 
accretion disks spectra: an application to Ultraluminous X-ray Sources}
\author[A. Lorenzin and L. Zampieri]{A. Lorenzin and L. Zampieri \\ %\thanks{E-mail: alessandro.lorenzin@yahoo.it}\\
INAF-Osservatorio Astronomico di Padova, Vicolo dell'Osservatorio 5, I-35122 Padova, Italy}
\begin{document}

\date{Accepted ... Received ...; in original form ...}
%
%\pagerange{\pageref{firstpage}--\pageref{lastpage}} \pubyear{2002}
%
\maketitle
\label{firstpage}
%
%
%
%%%%%%%%%%%%%%%%%%%%%%%%%%%%%%%%%%%%%%%%%%%%%%%%%%%%%%%%%%%%%%%%%%%%%%%%%%%%%
%
%
%
%
\begin{abstract}
We compare standard models of accretion  disks around black holes that
include the   appropriate  zero-torque  inner boundary   condition and
relativistic effects on   the emission and  propagation of  radiation.
The  comparison is  performed adopting  the multicolor disk  blackbody
model (MCD) as reference and looking for  the parameter space in which
it  is in statistical agreement  with ``more physical'' accretion disk
models.  We find simple 'recipes'  that can be  used for adjusting the
estimates of  the physical inner  radius  of the disk, the  black hole
mass and  the accretion rate inferred  using the parameters of the MCD
fits.  We applied  these results to four  ULXs for which MCD  spectral
fits of their  X-ray soft spectral  components have been published and
find that, in three cases (NGC 1313 X-1, X-2 and M 81 X-9), the black 
hole masses inferred for a standard disk around a Schwarzschild black 
hole are in the interval $\sim 100-200 M_\odot$.
Only if the black hole is maximally rotating are the masses comparable 
to the much larger values previously derived in the literature.
\end{abstract}
%
%
%
%%%%%%%%%%%%%%%%%%%%%%%%%%%%%%%%%%%%%%%%%%%%%%%%%%%%%%%%%%%%%%%%%%%%%%%%%%%%%
%
%
%
%
\section{Introduction}

Studying the spectra of X-ray  binaries is one  of the main tools  for
understanding the physics of these source and determine the properties
of the acccretion disk and the compact object  that they host. From an
observational standpoint,  Galatic  black hole candidates (BHCs)  show
a number of spectral states, defined in terms of  the components in their
X-ray spectra and the flux  level (e.g. \citealt{mr06,rm06}). 
In  the so called  \emph{high state} (\citealt{tan72,oda77}),
the spectrum is dominated by  a soft, thermal 
component up to 1-2 keV, with   a  power-law emerging at higher  energies.
In the \emph{low state} (\citealt{oda71,tan72})
the soft component is not present and the spectrum is
well   fitted  by a power-law  with   a cut-off at  few  tens  of keV.
Transitions to an \emph{intermediate state} (\citealt{men97,bel97})
or a \emph{very high state} (\citealt{miy91})
are also observed in which both components  are  present and equally 
important   in  terms of the emitted  flux. It  is worth  mentioning  
that this  classification is based  also on the timing  properties of the  
sources (e.g. \citealt{men98}). It is largely
accepted that the soft, thermal component originates from an accretion
disk, while the power-law is likely  produced by Comptonization of the
soft disk photons  from a hotter phase of  the accretion flow (usually
referred to as corona).  In  these  assumptions,  in  principle 
the properties of the disk  and the physical  parameters of the  accreting
black hole (BH) can be inferred from the study of the soft component.

Similar spectral states  (e.g.  \citealt{b23}; \citealt{b21})  and correlated  spectral/flux variability (e.g. \citealt{b38a}; 
\citealt{b50}) are also observed in Ultraluminous
X-ray   Sources (ULXs), very bright,   point-like X-ray sources in the
field of nearby galaxies  (see, e.g., \citealt{b19}). In fact, several
pieces   of evidence indicate  that the   majority  of ULXs  are X-ray
binaries, chiefly among them  the detection of periodic modulations in
the   X-ray  flux (e.g.    \citealt{b30b};    \citealt{f06})  and   the
identification         of      stellar    optical         counterparts
(e.g. \citealt{b40,b40a};  \citealt{b30a}; \citealt{b49}). The spectra
of several    ULXs show a   soft, thermal  component,  similar to that
observed  in  BHCs, which has been   interpreted  as emission  from an
accretion disk,  although for some  ULXs with spectral curvature above
2-3 keV different spectral models have also been proposed (e.g.
\citealt{b61}; \citealt{b27}; \citealt{m07}).

To model the soft component in BHCs and  ULXs the so called multicolor
disk blackbody  model (MCD; \citealt{b48})    has been often   adopted
because it  is  easy to  use and is   efficiently implementated in the
X-ray spectral fitting  package XSPEC.  The  fitting parameters of the
MCD model  (the disk  inner temperature and  the normalization  of the
spectrum) depend   on  the BH mass   and  accretion rate,  and  can in
principle be used to estimate the physical parameters of the accretion
flow and   of the BH.   However,  it  is well known  that  this  model
represents  an approximation  of   the standard,  Shakura  \& Syunyaev
(1973) disk and differs from it for  the expression of the temperature
profile, in which   the effects of  the  viscous torque at   the inner
boundary are neglected.  Also, the locally emitted spectrum is assumed
to be  a blackbody, neglecting  the vertical temperature structure and
atmospheric radiative transfer  effects.   Although the effects  of  a
zero-torque     inner   boundary  in   a  pseudo-Newtonian   potential
(\citealt{b54a})    were  implemented   in   the \emph{diskpn}   model
(\citealt{b25}), temperature  profile correction  factors for the  MCD
model were also derived  (e.g.  \citealt{b36,m00}).  At the same time,
limited deviations from LTE  caused by radiative transfer effects were
accounted for using  an  hardening  factor $f=T_{col}/T_{eff}$,  where
$T_{col}$ and $T_{eff}$ are the color and effective temperatures (e.g.
\citealt{b59,b73a}).   Even  adopting all these correction factors,
deviations of the MCD with respect to a  standard disk spectrum may be
significant if the physical state of  the disk changes or the coupling
bewteen the disk  and the corona   becomes important, which  may occur
especially at high accretion rates (\citealt{b43}).

Despite all  these uncertainties, the parameters of  the MCD  fit have
been used for estimating the mass of the  BH and the accretion rate in
X-ray binary  systems.  This has recently  been done also for ULXs for
which,  at  present, X-ray   spectroscopy   provides one  of   the few
available methods to  obtain zero-th order estimates  of the  BH mass.
Indeed, if the soft component observed in ULX spectra is fitted with a
MCD model (e.g. \citealt{b44,b44a}), its characteristic temperature is
typically  $\sim 200$ eV,  much lower  than  that observed in Galactic
BHCs.  Both the low temperature and high normalization constant of the
MCD   model  have been  used   to  estimate the  BH  masses, typically
obtaining values  largely in excess  of 100 $M_\odot$.   Modelling the
soft component in  these terms is clearly possible  if ULXs are  in an
accretion stage similar to  that of BHCs  in the soft (or nearly soft)
state, so that  a disk spectral  component is present.  Along with the
very high     (isotropic)  luminosity    of   these  sources,    these
``spectroscopic  estimates'' have led  to the suggestion that they may
contain intermediate mass black holes (IMBHs).

In this  paper  we revisit and compare   standard models  of accretion
disks  around black holes  that   include the appropriate  zero-torque
inner boundary condition and/or relativistic effects  on the emission and
propagation of radiation.  They  are derived in the  usual assumptions
that:  (1) the accretion disk  is  optically thick, geometrically thin
and in  a quasi-steady state; (2) the  locally emitted spectrum from a
small annulus of  the disk is  a blackbody ($I_{\nu}=B_{\nu}$), with a
color correction  factor  $f$  taking  into  account  for  atmospheric
transfer effects.  The comparison  is  performed adopting the  MCD  as
reference model and looking for the  parameter space in which it turns
out  to be in agreement  with ``more physical'' accretion disk models.
The fits  are performed keeping  either the BH   mass or the accretion
rate  fixed and allowing the inner  radius of the MCD   to vary.  This
gives us simple 'recipes' that can be used for adjusting the estimates
of the physical inner radius of the disk, the black  hole mass and the
accretion rate inferred using the parameters of the MCD fits.  
%of the soft component in BHCs and ULXs spectra.

The paper is organized as follows.  Section 2 contains a short summary
of    the basic  structure  equations  of  the  accretion  disk models
considered   in  this work;  Section  3  presents  the  results of the
comparison among  the different spectral models;  Section 4 is devoted
to  an  application  of  the  results of    this  comparison to   some
representative   ULX  spectra,   while    Section  5   summarizes  our
conclusions.

%
%
%
%
%%%%%%%%%%%%%%%%%%%%%%%%%%%%%%%%%%%%%%%%%%%%%%%%%%%%%%%%%%%%%%%%%%%%%%%%%%%%%
%
%
%
%
\section{Accretion disk spectra}
\label{Accretion disk spectra}

In this section we summarize the basic equations of the accretion disk
models that we have considered in our work. We start with recalling
the main properties of the standard disk model \citep{b57} and the
multicolor disk blackbody model (MCD, \citealt{b48}), that are usually
adopted in fitting the spectra of X-ray binaries. Then, we shortly
review the structure equations for a relativistic accretion disk
orbiting around a Schwarzschild and a Kerr black hole
\citep{b4,b53,b54}. The emergent spectrum is calculated taking into
account the relativistic effects on the emission and propagation of
light (gravitational redshift, light bending, Doppler shift) using the
formalism of the transfer function introduced by \citet{b14}. More
sophisticated treatments of photon propagation in curved space-times,
both numerical \citep{b37,b31,b6,b20,b71,b56,b51,b5} and also
analytical \citep{b7,b26,b8,b10} are available. Such techniques adopt
ray tracing of the photon geodetics from the disk plane to the
observer and are particularly suited when dealing with the calculation
of spectral line profiles. However, we are essentially interested in
comparing the continuum spectral shape of the MCD disk with that of a
standard and relativistic disk and, to this end, a treatment of the
relativistic geometrical optics a l\`a \citet{b14} appears acceptable.
%
%
%
%
%%%%%%%%%%%%%%%%%%%%%%%%%%%%%%%%%%%%%%%%%%%%%%%%%%%%%%%%%%%%%%%%%%%%%%%%%%%%%
%
%
%
%
\subsection{Standard accretion disk and MCD}
\label{Standard accretion disk and MCD}

Let's consider the \emph{Standard accretion disk} model \citep{b57} in
the usual approximations (Newtonian, optically thick, geometrically
thin and in a quasi-steady state). The expression of the temperature
profile is obtained setting the energy rate of viscous dissipation
equal to the total radiation flux (e.g. \citealt{b24}):
\begin{equation}
T\left(r\right)=T_{\ast}\left(\frac{r}{R_{in}}\right)^{-3/4}
\left[1-\left(\frac{R_{in}}{r}\right)^{1/2}\right]^{1/4},\\ \label{A0}
\end{equation}
where
\begin{equation}
T_{\ast}=\left( \frac{3GM\dot{M}}{8\pi\sigma R_{in}^{3}} \right)^{1/4} \, , \\
\label{A01}
\end{equation}
$M$ is   the  black hole mass,   $\dot{M}$ is  the  accretion rate and
$R_{in}$ is the inner disk  radius. The emitted spectrum is calculated
assuming that the disk emits locally as a blackbody, so that:
\begin{equation}
L_\nu = 4\pi r_0^2 F_{\nu}=8\pi^2 \cos i \int^{R_{out}}_{R_{in}}I_{\nu}(T)rdr,\\ \label{A1}
\end{equation}
where $i$ is the disk's inclination angle with respect to the
direction of the line of sight, $r_0$ is the distance to the observer,
$R_{out}$ is the outer disk radius and $I_{\nu}(T)$ is the specific
intensity. We assume that the gas is in LTE and hence
$I_{\nu}(T)=B_{\nu}(T)$, where $B_{\nu}(T)$ is the Planck function.
%It is known that, for high accretion rates, in the inner part of the
%disk radiation pressure and electron scattering become important and
%it is no longer possible to assume $I_{\nu}=B_{\nu}(T)$ (see e.g.
%\citealt{b43}).

The MCD model is an approximation of the standard disk and differs
from it only for the expression of the temperature profile, in which
the effects of the viscous torque at the inner boundary are
neglected. The temperature profile is then given by \citep{b48}:
\begin{equation}
T\left(r\right)=T_{in}\left(\frac{r}{R_{in,BB}}\right)^{-3/4},\\ \label{A2}
\end{equation}
where
\begin{equation}
T_{in}=\left(\frac{3GM\dot{M}}{8\pi\sigma R_{in,BB}^3}\right)^{1/4} \, , \\ \label{A02}
\end{equation}
and $R_{in,BB}$ is the inner radius of the MCD. The emitted spectrum
can again be calculated from equation~(\ref{A1}), with the
temperature profile given by equation~(\ref{A2}).

In the following, we will use the MCD model as reference for fitting
the standard and relativistic disk spectra, because this model is
frequently adopted for the spectral fits of X-ray binaries.  The
fitting parameters of the MCD model, the disk inner temperature
$T_{in}$ and the spectral normalization factor $K_{BB}=\cos i
(R_{in,BB}/r_0)^2$, depend on $M$ and ${\dot M}$. Assuming that the
inner radius $R_{in,BB}=b R_g$ ($R_g=GM/c^2$ is the gravitational
radius), it is possible to write $M$ and ${\dot M}$ as:
\begin{eqnarray}
&& \frac{M}{M_\odot}=\frac{67.5}{b} \left(\frac{r_0}{1 \, {\rm Mpc}}\right) \left(\frac{K_{BB}}{\cos i}\right)^{1/2} \label{A3} \\
&& \frac{\dot{M}}{\dot M_{\rm Edd}}=0.1 b^2 \left(\frac{r_0}{1 \, {\rm Mpc}}\right) \left(\frac{K_{BB}}{\cos i}\right)^{1/2} \left(\frac{T_{in}}{\rm 1 keV}\right)^4
\, , 
\label{A4}
\end{eqnarray}
where
$K_{BB}=\left[\left(R_{in,BB}/1 \, {\rm Km}\right)^{2}\left(r_0/10 \, {\rm kpc}\right)^{-2}\cdot\cos
i\right]$. 

In these assumptions, if the distance $r_0$ is known and the inclination
angle $i$ is fixed, the MCD parameters can be used for estimating $M$
and/or ${\dot M}$. These expressions depend on the value of $b$, the
inner radius in units of $R_g$, and $T_{in}$. It is well known that these estimates
of the BH mass and the disk accretion rate are uncertain, because the
physical state of the disk may vary and, in addition, radiative
transfer effects induced by scattering may cause departures of
$I_{\nu}$ from $B_{\nu}$. 
Several non-LTE accretion disk models, in which radiative transfer 
effects at the surface of the disk are computed, have been
presented in the literature, for both hot disks around stellar mass BHs
(e.g. \citealt{b43,b14a}) and cool disks around intermediate mass
BHs (e.g. \citealt{b36d}).
These effects are very important in predicting the local spectra emitted 
at a given disk annulus and can be accounted for using an
hardening factor $f=T_{col}/T_{eff}$, where $T_{col}$ and $T_{eff}$
are the color and effective temperature. It is easy to show that
$K_{BB}\propto 1/f^4$ and, as a consequence, $M\propto f^{2}$ and
${\dot M}/{\dot M_{Edd}}\propto f^{2}$ (\citealt{b36,b36a}). So, as a
first approximation, we account for radiative transfer effects simply
inserting $f^{2}$ into equations (\ref{A3}) and
(\ref{A4}). Computations of $f$ for standard and relativistic disks
give values in the interval 1.4-2 (see \citealt{b59}; \citealt{b73a};
\citealt{b36d,b14a}).

%
%
%
%
%%%%%%%%%%%%%%%%%%%%%%%%%%%%%%%%%%%%%%%%%%%%%%%%%%%%%%%%%%%%%%%%%%%%%%%%%%%%%
%
%
%
%
\subsection{Relativistic accretion disk}
\label{Spectrum of a relativistic accretion disk}

The structure equations and the emitted spectrum of the relativistic
disk are obtained in the same assumptions stated above: (1) the
accretion disk is optically thick, geometrically thin and in a
quasi-steady state; (2) the local emitted spectrum from a small
annulus of the disk is a blackbody ($I_{\nu}=B_{\nu}$).

In the following, we will refer to the radius of the innermost stable
circular orbit around a black hole with $r_{ms}$. Its expression is
given by:
\begin{equation}
r_{ms}=M\left\{3+Z_{2}-\left[\left(3-Z_{1}\right)\left(3+Z_{1}+2Z_{2}\right)\right]^{1/2}\right\},\\ \label{A5}
\end{equation}
where $Z_1$ and $Z_2$ are functions of $M$ and $a$ \citep{b53}, and
$a$ is the specific angular momentum of the black hole. 
\subsubsection{Temperature profile of the relativistic disk}
\label{Relativistic disk structure}

A detailed derivation of the structure of a relativistic accretion
disk around a black hole is presented in \citet{b53} and \citet{b54}
to which we refer for all the details. Here we present only the relevant
equations for the temperature profile and the locally emitted
flux.

Following \citet{b53}, the emitted radiation flux can be expressed as
a function of the integrated $\phi-r$ component of the stress-energy
tensor (evaluated in the comoving frame of the observer in geodetic
circular motion) as:
\begin{equation}
%F=\frac{3\dot{M}}{8\pi r^{2}}\frac{M^{-2}}{r}\frac{\emph{Q}}{\emph{B}\emph{C}^{1/2}},\\ \label{A12}
F=\frac{3GM\dot{M}}{8\pi r^{3}}\frac{\emph{Q}}{\emph{B}\emph{C}^{1/2}},\\ \label{A12}
\end{equation}
where $B$, $C$ and $Q$ depend on the radial 
coordinate $r$ and the specific angular momentum of the 
black hole $a$ (\citealt{b53}). The function $Q$ depends also on
the angular momentum per unit mass of a generic circular orbit and of
the innermost stable circular orbit (\citealt{s76}).
Assuming that the radiative transport of energy is dominant over the
turbolent one, in LTE we can write
\begin{equation}
F=\sigma T^{4}.\\ \label{A13}
\end{equation} 
Comparing equations (\ref{A12}) and (\ref{A13}) it is possible to obtain the 
surface temperature profile for a relativistic accretion disk:
\begin{equation}
%T_{rel}(r)=3\cdot10^{7}m^{-1/2}_{\ast}\dot{m}^{1/4}_{0\ast}r^{-3/4}\emph{B}^{-1/4}\emph{C}^{-1/8}\emph{Q}^{1/4},\\ \label{A14}
T(r)=T_{\ast}
\left(\frac{r}{R_{in}}\right)^{-3/4}\emph{B}^{-1/4}\emph{C}^{-1/8}\emph{Q}^{1/4},\\
\label{A14} \, ,
\end{equation}
where $T_{\ast}$ is given by equation~(\ref{A01}).
\subsubsection{Relativistic effects on the propagation of radiation}
\label{Relativistic effects on the propagation of radiation}

Following \cite{b4} and \cite{b14} the spectrum of a relativistic
accretion disk is calculated by means of a trunsfer function $f$, that
accounts for all relativistic effects on the emission and propagation
of radiation (Doppler shift, light bending and gravitational
redshift). The expression for $f$ is \citep{b14}:
\begin{equation}
f\left(g^{\ast},r_{e},\vartheta_{0}\right)dg^{\ast}dr_{e}=\frac{g}{\pi
r_{e}}
\left(g^{\ast}-g^{\ast^{2}}\right)^{1/2}r^{2}_{0}\cos \vartheta_{0}d\Omega_{0},\\ \label{A15}
\end{equation}
where $r_{e}$ is the radius of a generic emitting ring of the disk,
$\vartheta_{0}$ is the polar angle between the line of sight of the
distant observer and the disk's polar axis, $d\Omega_{0}$ is the solid
angle that a specific geodesics family subtends to the observer at a
given energy and for a given emission radius, $g=E_{0}/E_{e}$ is the
ratio of the observed photon energy and the emitted one, $g^{\ast}$,
$g_{max}$ and $g_{min}$ are functions of $g$ (for their expression see
\citealt{b14}).

The radiation flux seen by a distant observer is:
\begin{equation}
F_{0}=\int I_{0}\cos \vartheta_{0}d\Omega_0,\\ \label{A16}
\end{equation}
where $I_{0}$ is the observed specific intensity. If $I_{e}$ is the
emitted specific intensity, from the expression of the relativistic
invariant $I/E^3$ 
and equation~(\ref{A16}) 
we obtain the observed luminosity per unit energy
\begin{equation}
L_0=4\pi r^{2}_{0}F_{0}=4\pi\int g^{3}I_{e}r^{2}_{0}\cos \vartheta_{0}d\Omega_0.\\ \label{A19}
\end{equation}
Using equation~(\ref{A15}) it is possible to rewrite $r^{2}_{0}\cos
\vartheta_{0}d\Omega_0$ in terms of
$f\left(g^{\ast},r_{e},\vartheta_{0}\right) dg^* dr_e$. Inserting the resulting
expression into equation (\ref{A19}), we obtain:
\begin{equation}
L_{0}=\int 2\pi f\ g^{2}I_{e}\left(g^{\ast}-g^{\ast^{2}}\right)^{-1/2}
dg^{\ast}d\left(\pi r^2_{e}\right).\\ \label{A20}
\end{equation}
Tabulated values of the transfer function for a Schwarzschild and a maximally
rotating ($a=0.9981$) Kerr BH were reported by \cite{b14}. In the following we will 
consider these two extreme cases as reference.
%
%
%
%
%%%%%%%%%%%%%%%%%%%%%%%%%%%%%%%%%%%%%%%%%%%%%%%%%%%%%%%%%%%%%%%%%%%%%%%%%%%%%
%
%
%
%
\begin{figure}
 \includegraphics[width=84mm]{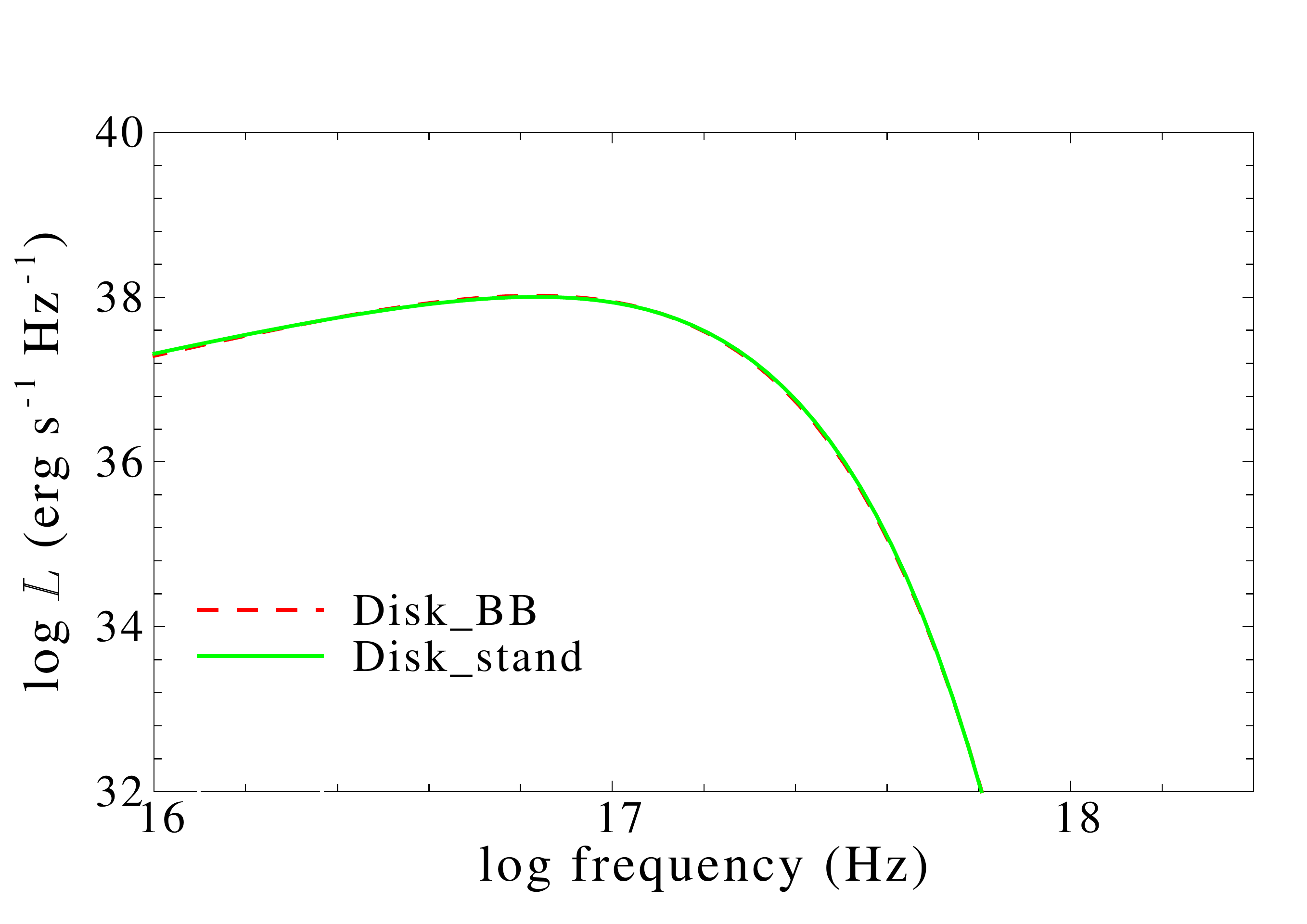}
 \includegraphics[width=84mm]{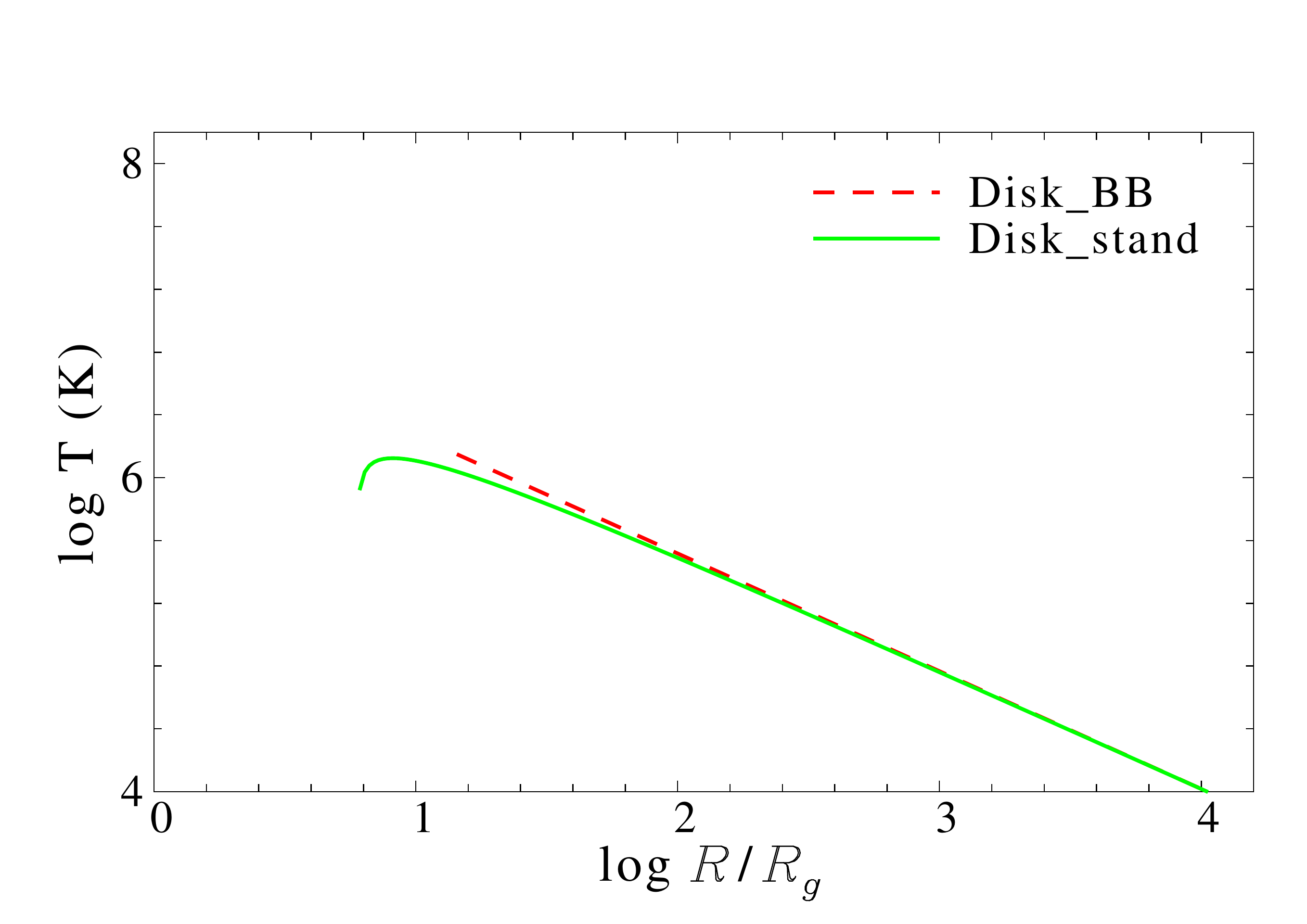}
 \caption{Spectrum ({\it top panel}) and temperature profile {\it bottom panel}) of the \emph{Disk-stand} model ({\it solid-green}) 
 for $M_{stand}=100M_{\odot}$, $\dot{M}_{stand}=0.1\dot{M}_{Edd}$, $R_{in}=6R_{g}$, and the corresponding best-fitting
 \emph{Disk-BB} model ({\it dashed-red}; $M_{BB}=94M_{\odot}$, $\dot{M}_{BB}=0.09\dot{M}_{Edd}$, $R_{in,BB}=13.9R_{g}$).
 These plots refer to the fitting results reported in Table~\ref{tab0}, second row ($\cos i=1$).
 }
 \label{figure1}
\end{figure}

\section{Parameter estimates from spectral fits}
\label{Parameter estimates from spectral fits}

We implemented the structure equations and spectra of the accretion
disk models outlined above and produced fits of the standard and
relativistic disks against the MCD model.
This model-model comparison has been performed by means of a standard
$\chi^2$ minimization procedure, that provides also an assessment of
the statistical significance of the fit. A fixed reference error of
$\sim$10\% was assigned to the spectra. As mentioned above, we use the
MCD model as reference for fitting all the other disk spectra, because
this model is frequently adopted for the spectral fits of X-ray
binaries within XSPEC. The values of the best-fitting MCD parameters
are then compared with those of the other models.

From now on we refer to the different accretion disk models with the
following expressions: \emph{Disk-stand} for the standard accretion
disk \citep{b57}, \emph{Disk-BB} for the MCD \citep{b48},
\emph{Disk-rel} or \emph{Disk-kerr} for a relativistic disk around a
Schwarzschild or a maximally rotating ($a=9981$) Kerr black hole \citep{b53}, respectively. We will
also use different indeces to denote the mass $M$ and the accretion
rate $\dot{M}$ of the different models ($M_{BB}$, $\dot{M}_{BB}$ for
the \emph{Disk-BB}; $M_{stand}$, $\dot{M}_{stand}$ for the
\emph{Disk-stand}; $M_{rel}$, $\dot{M}_{rel}$ for the
\emph{Disk-rel}; $M_{kerr}$, $\dot{M}_{kerr}$ for the
\emph{Disk-kerr}). The accretion rate is expressed in units of the 
Eddington accretion rate ${\dot M_{Edd}}=L_{edd}/c^2$, where
$L_{edd}$ is the Eddington luminosity. For an accretion efficiency
$\eta = 0.1-0.4$, the accretion luminosity reaches $L_{Edd}$ for 
${\dot M} = {\dot M_{Edd}}/\eta = 2.5-10 {\dot M_{Edd}}$.

\subsection{\emph{Disk-stand} model}

Fits of the standard disk with a MCD were performed in the literature
in order to obtain correction factors for the inner disk radius, the
black hole mass and the accretion rate (e.g. \citealt{b36,m00}). We
performed similar fits of the \emph{Disk-stand} with the
\emph{Disk-BB} but keeping either the BH mass or the accretion rate fixed.
Acceptable fits with different best-fit values of $R_{in,BB}$ were
obtained (see Table~$\ref{tab0}$). 
For the fit with fixed $M$, we obtain $R_{in,BB}=13R_{g}$, while for
the fit with fixed $\dot{M}$, $R_{in,BB}=14.9R_{g}$. In the first
case, the value of $\dot{M}_{BB}$ returned by the fit is $\approx
20\%$ smaller than $\dot{M}_{stand}$ while, in the second case, the
values of $M_{BB}$ derived from the fit is $\approx 12\%$ smaller than
the corresponding value of $M_{stand}$. Clearly, there are also fits
with fixed $R_{in,BB}$ in the interval 13$\leq R_{in,BB}\leq$14.9,
that return intermediate values of $M_{BB}$ and $\dot{M}_{BB}$ (see
again Table~$\ref{tab0}$). In Figure~$\ref{figure1}$ we show the
results of one of these fits.
%obtained fixing $R_{in,BB}\approx 13.9R_{g}$. In this case, the values
%of both $M_{BB}$ and $\dot{M}_{BB}$ inferred from the fit differ from
%those of the \emph{Disk-stand} model by $\sim 10-15\%$.

These results are in agreement with those of
\cite{b36}, who showed that the inner radius of the best-fitting MCD model
does not correspond to the true inner radius of the standard disk, but
that the two radii are related by the expression (see also
\citealt{b75})
\begin{equation}
R_{in,BB}=2.6 R_{in} \, .
\label{A21}
\end{equation}
In fact, in order for the spectral peaks of the \emph{Disk-BB} and
\emph{Disk-stand} be superimposed (see Figure $\ref{figure1}$), the
maximum temperature of the standard disk (0.488$T_*$; see
equations~[\ref{A0}] and [\ref{A01}]) must be equal to the maximum
temperature of the \emph{Disk-BB} ($T_{in}$; see
equation~[\ref{A02}]). From this, equation~(\ref{A21}) easily
follows. Therefore, assuming that the \emph{Disk-stand} terminates at
the ISCO (innermost stable circular orbit), $R_{in}=6 R_G$ and the
inner radius of the best-fitting \emph{Disk-BB} must be located at
$\simeq 15.6 R_g$, in subtantial agreement with the results of the model-model fits.

We note that all the results presented here hold for any inclination angle,
as the \emph{Disk-BB} and \emph{Disk-stand} models have the same dependence 
on $i$.
\begin{table*}
 \begin{center} \begin{minipage}{140mm} 
 \caption{Values of the characteristic parameters of the \emph{Disk-stand} 
 model and those derived by the spectral fits with the \emph{Disk-BB} model 
 for $\cos i=1$. The inner radius of the \emph{Disk-stand} model is $6 R_g$. 
 In the first row $M_{BB}$ is set equal to $M_{stand}$; in the second row 
 $R_{in,BB}$ is fixed at an intermediate value
 between 13 and 14.9 $R_g$; in the last row ${\dot M}_{BB}$ is set equal to 
 ${\dot M}_{stand}$. The results of the spectral fit at fixed radius are shown
 in Figure~\ref{figure1}.
 }
 \label{tab0}
 \begin{center}
 \begin{tabular}{@{}lcccccc}
  \hline
   &  $M_{stand}$ & ${\dot M}_{stand}$ & $M_{BB}$ & ${\dot M}_{BB}$ & $R_{in,BB}$ & $\chi^{2}_{red}$ \\
   & ($M_{\odot}$) & ($M_{\odot}$) & ($\dot{M}_{Edd}$) &  ($\dot{M}_{Edd}$) & ($R_{g}$) & \\    
  \hline
   & 100.0 & 0.1 & 100.0 & 0.08$\pm$0.01 & 13.0$\pm$1.5 & 0.1 \\
   
   & 100.0 & 0.1 & 94.0$\pm$10.0 & 0.09$\pm$0.03 & 13.9 & 0.2 \\
   
   & 100.0 & 0.1 & 88.0$\pm$32.0 & 0.1 & 14.9$\pm$1.3 & 0.2 \\
  \hline
  \end{tabular}
  \end{center}
  \medskip
  
  \end{minipage}
  \end{center}
\end{table*}
\begin{table*}
 \begin{center}
 \begin{minipage}{140mm}
 \caption{Values of the characteristic parameters of the \emph{Disk-rel} 
 model and those derived by the spectral fits with the \emph{Disk-BB} 
 model for $\cos i=1$ and $\cos i=0.5$. The inner radius of the 
 \emph{Disk-rel} is $R_{in,rel}=6R_{g}$. 
 In the first row $M_{BB}$ is 
 set equal to $M_{rel}$; in the second row $R_{in,BB}$ is fixed at an intermediate
 value between 19.2 and 25 $R_g$; in 
 the last row ${\dot M}_{BB}$ is set equal to ${\dot M}_{rel}$ (for 
 each value of $\cos i$). The results of the spectral fit at fixed radius are shown
 in Figure~\ref{figure2}.
 }
 \label{tab1}
 \begin{center}
 \begin{tabular}{@{}lcccccc}
  \hline
   &  $M_{rel}$ & ${\dot M}_{rel}$ & $M_{BB}$ & ${\dot M}_{BB}$ & $R_{in,BB}$ & $\chi^{2}_{red}$ \\
   & ($M_{\odot}$) & ($M_{\odot}$) & ($\dot{M}_{Edd}$) &  ($\dot{M}_{Edd}$) & ($R_{g}$) & \\
  \hline     
   $\cos i=1$ & 50.0 & 0.1 & 50.0 & 0.06$\pm$0.03 & 19.2$\pm$11.6 & 0.2 \\
   
    & 50.0 & 0.1 & 43.5$\pm$6.7 & 0.07$\pm$0.02 & 21.6 & 0.1 \\
   
    & 50.0 & 0.1 & 37.3$\pm$10.2 & 0.1 &  25.0$\pm$7.1 & 0.1 \\
  \hline
   $\cos i=1$ & 100.0 & 0.1 & 100.0 & 0.06$\pm$0.03 & 19.2$\pm$3.5 & 0.2 \\
   
    & 100.0 & 0.1 & 87.0$\pm$31.0 & 0.07$\pm$0.02 & 21.6 & 0.1 \\
   
    & 100.0 & 0.1 & 75.0$\pm$34.0 & 0.1 &  25.0$\pm$6.7 & 0.1 \\
  \hline
  \hline
   $\cos i=0.5$ & 100.0 & 0.1 & 100.0 & 0.03$\pm$0.05 & 11.2$\pm$3.4 & 1.3 \\
   
    & 100.0 & 0.1 & 50.0$\pm$32.0 & 0.1 & 21.5$\pm$12.3 & 1.8 \\
  \hline
  \end{tabular}
  \end{center}
  \medskip
  
  \end{minipage}
  \end{center}
\end{table*}
\begin{figure}
 \includegraphics[width=84mm]{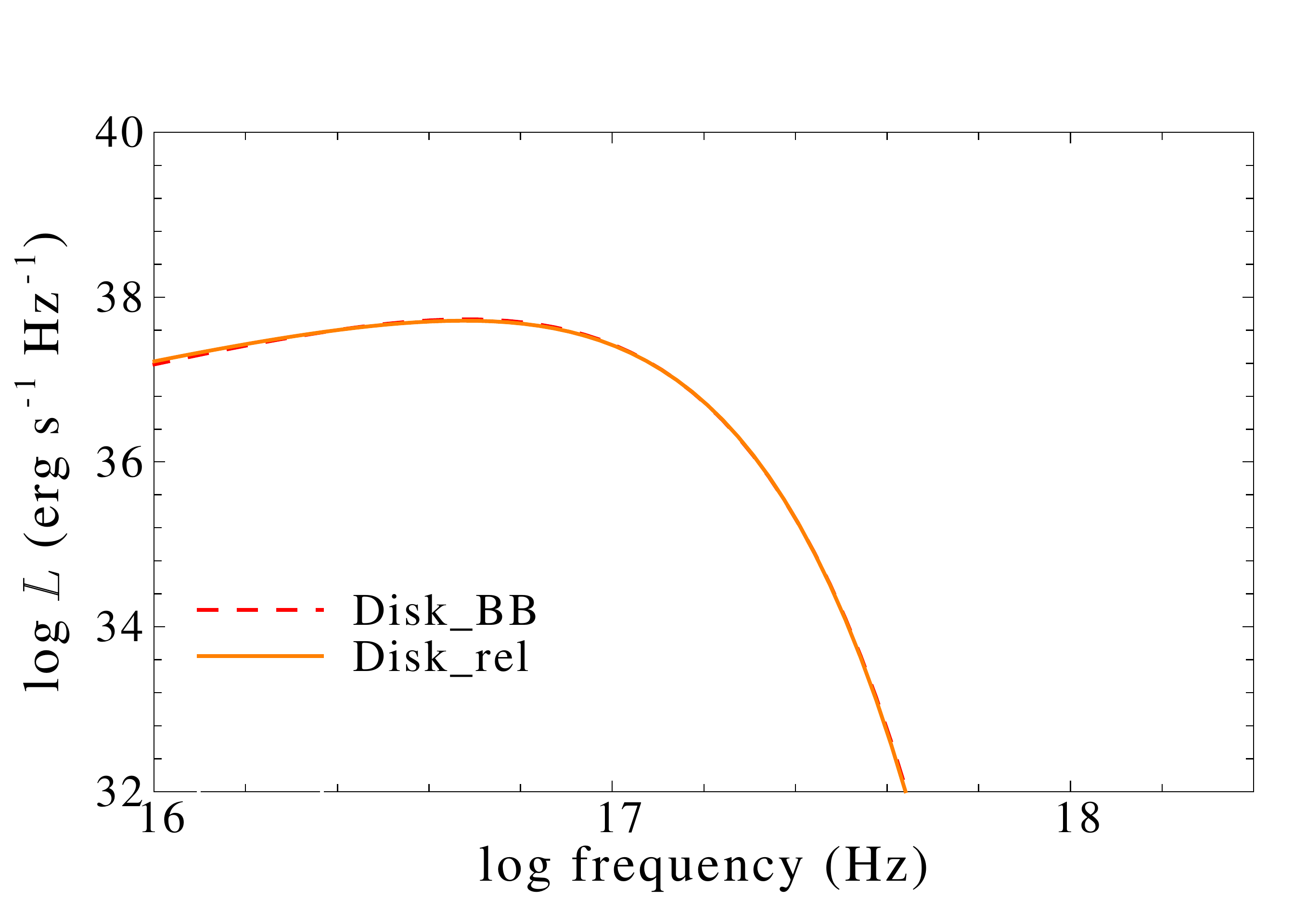}
 \includegraphics[width=84mm]{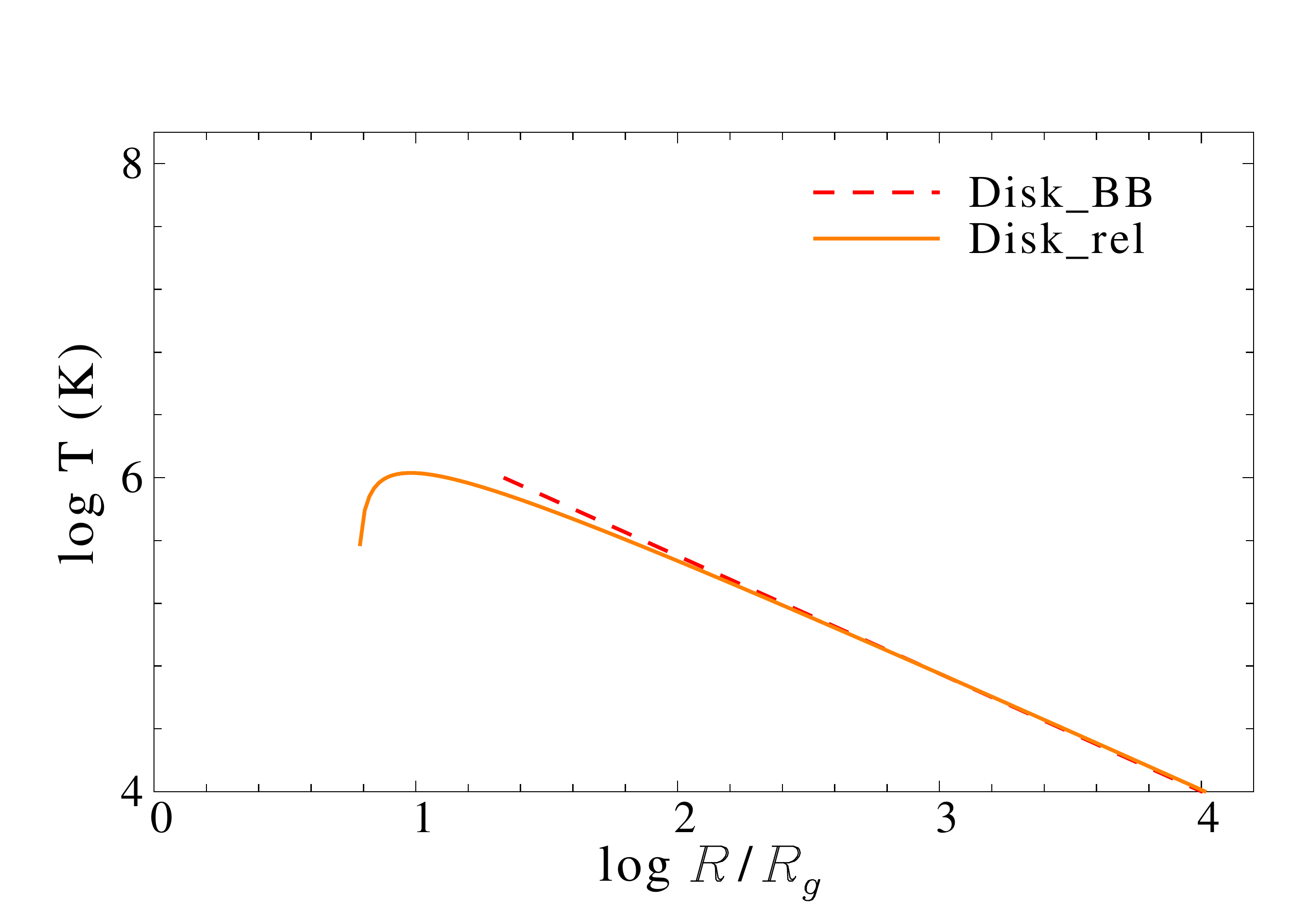}
 \caption{Same as Figure~\ref{figure1} for the \emph{Disk-rel} model ({\it solid-orange}; $M_{rel}=100M_{\odot}$, 
 $\dot{M}_{rel}=0.1\dot{M}_{Edd}$, $R_{in,rel}=6R_{g}$). The parameters of the best-fitting \emph{Disk-BB} model ({\it dashed-red})
 are $M_{BB}=87M_{\odot}$, $\dot{M}_{BB}=0.07\dot{M}_{Edd}$, $R_{in,BB}=21.6R_{g}$
 (see Table~\ref{tab1}, fifth row). For both models $\cos i=1$.}
 \label{figure2}
\end{figure}
\subsection{\emph{Disk-rel} model}

Similar fits of the \emph{Disk-rel} model with the \emph{Disk-BB} were
also performed.  For sufficiently small inclination angles, also in
this case the emitted spectrum can be well reproduced by the MCD.
The fits were performed keeping either $M$ or ${\dot M}$ fixed.
Results are shown in Table~$\ref{tab1}$. The two fits return different
values of the inner radius, $R_{in,BB}=19.2R_{g}$ for fixed $M$ and
$R_{in,BB}=25R_{g}$ for fixed $\dot{M}$. In the first case the value
of $\dot{M}_{BB}$ returned by the fit is $\approx 40\%$ smaller than
$\dot{M}_{rel}$ wheras, in the second case, the value of $M_{BB}$
derived from the fit is $\approx25\%$ smaller than the corresponding
value of $M_{rel}$. From Table~$\ref{tab1}$ we can see that good fits
can be obtained also fixing $R_{in,BB}$ in the interval 19.2--25$R_{g}$ and
leaving $M_{BB}$ and $\dot{M}_{BB}$ free to vary. In
Figure~$\ref{figure2}$ we plot the results of one of these fits
($R_{in,BB}\approx 21.6R_{g}$).
 
If we increase the inclination angle, the relativistic effects
(especially the gravitational focusing) tend to increase the
contribution of radiation coming from the inner parts of the disk,
making the spectrum of the \emph{Disk-rel} intrinsically harder than
that of the \emph{Disk-BB}. In fact, for high inclination angles the
\emph{Disk-BB} is only in rough agreement (within a $\sim 10 \%$
fractional error) with the \emph{Disk-rel} spectrum, and no
satisfactory fit is obtained ($\chi^2_{red} \geq 1.3$). Similar results
for $\cos i<1$ were obtained by \citet{b15}. For illustrative purposes
alone, in Table~$\ref{tab1}$ we report also the values of the fits for
$\cos i=0.5$.
\begin{figure}
 \includegraphics[width=84mm]{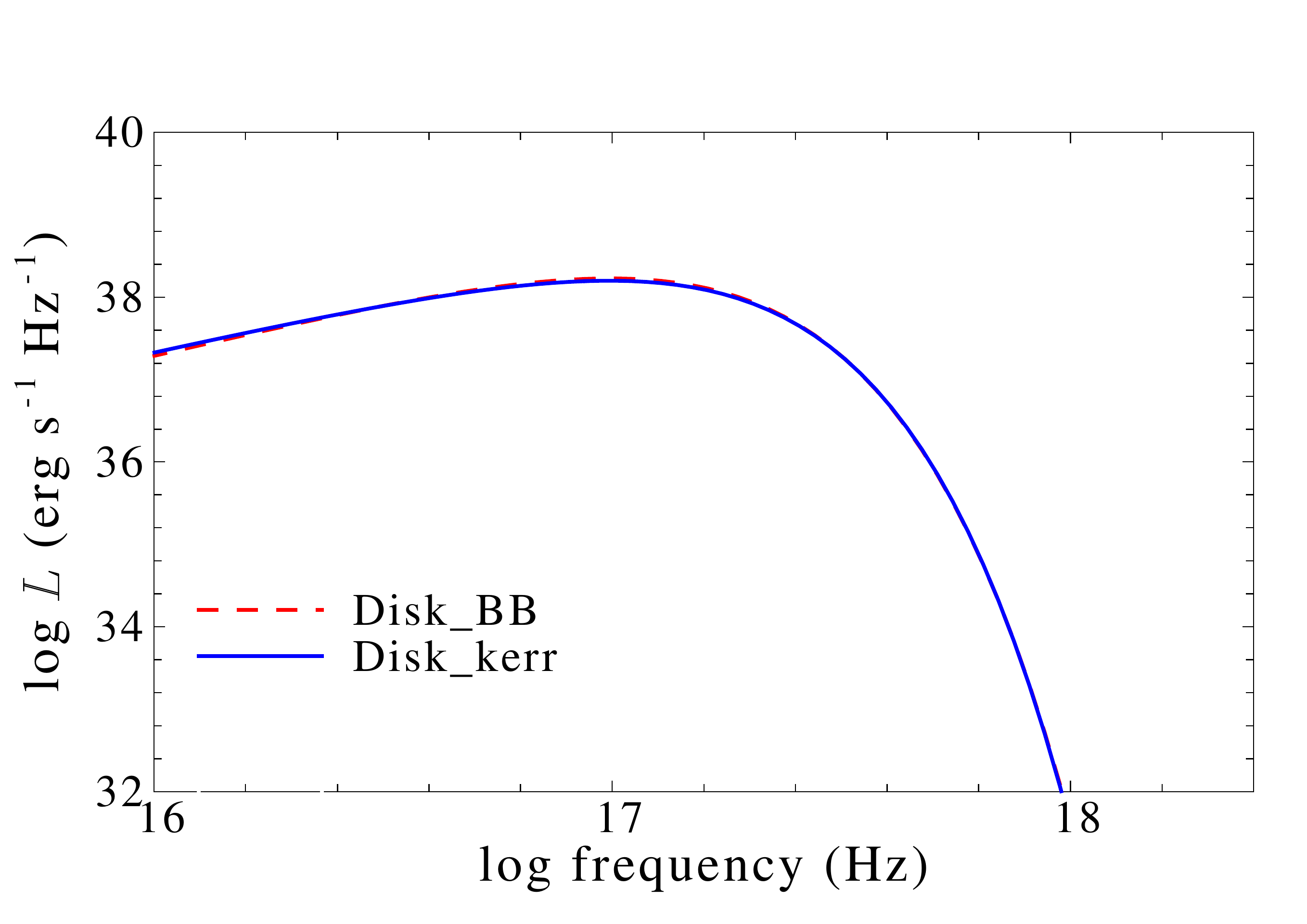}
 \includegraphics[width=84mm]{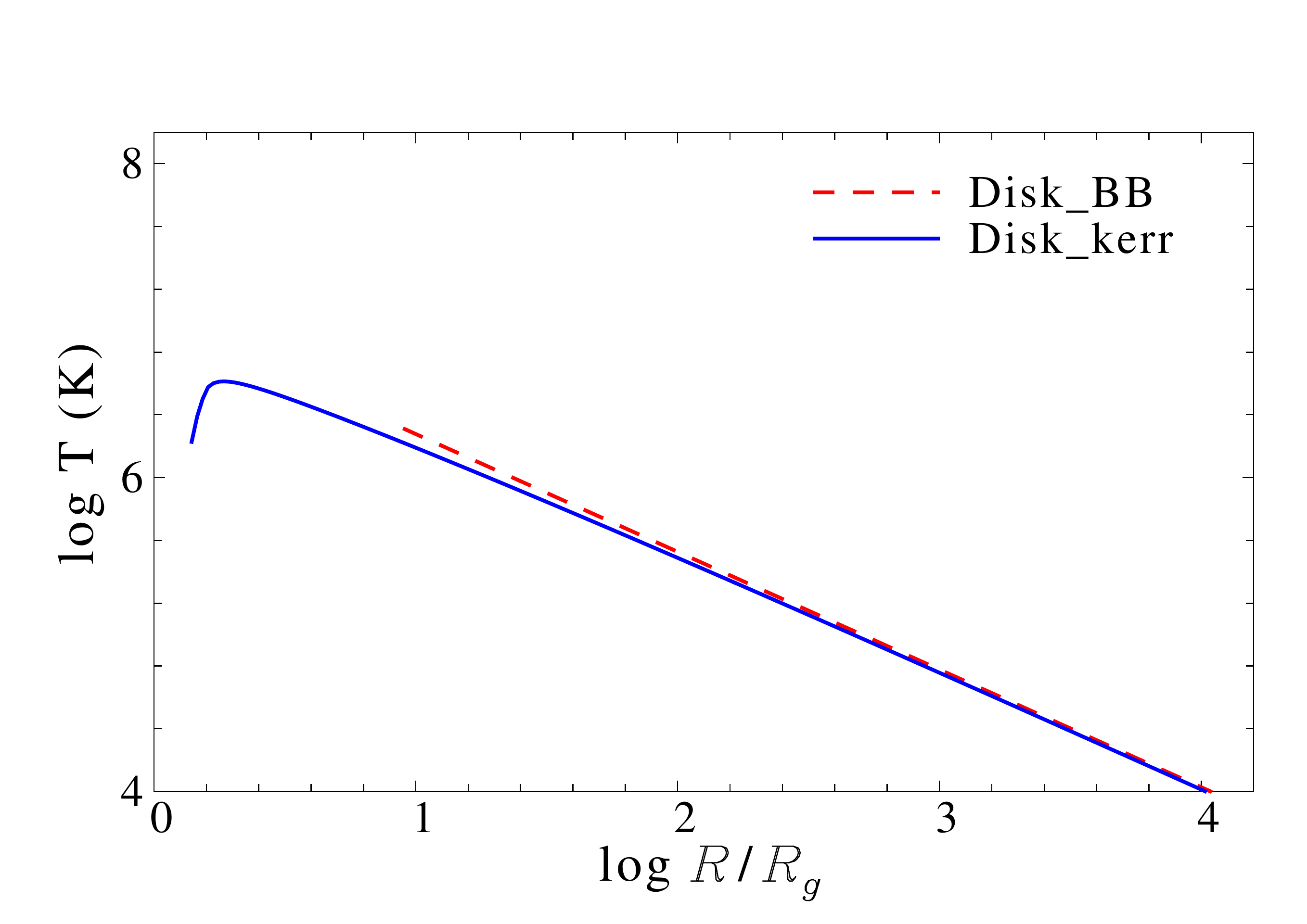}
 \caption{Same as Figure~\ref{figure1} for the \emph{Disk-kerr} model ({\it solid-blue}; $M_{kerr}=100M_{\odot}$, $\dot{M}_{kerr}=0.1\dot{M}_{Edd}$). 
 The parameters of the best-fitting \emph{Disk-BB} model ({\it dashed-red})
 are $M_{BB}=88M_{\odot}$, $\dot{M}_{BB}=0.095\dot{M}_{Edd}$, $R_{in,BB}=8.9R_{g}$
 (see Table~\ref{tab2}, second row). For both models $\cos i=1$.}
 \label{figure3}
\end{figure}
\begin{figure}
 \includegraphics[width=84mm]{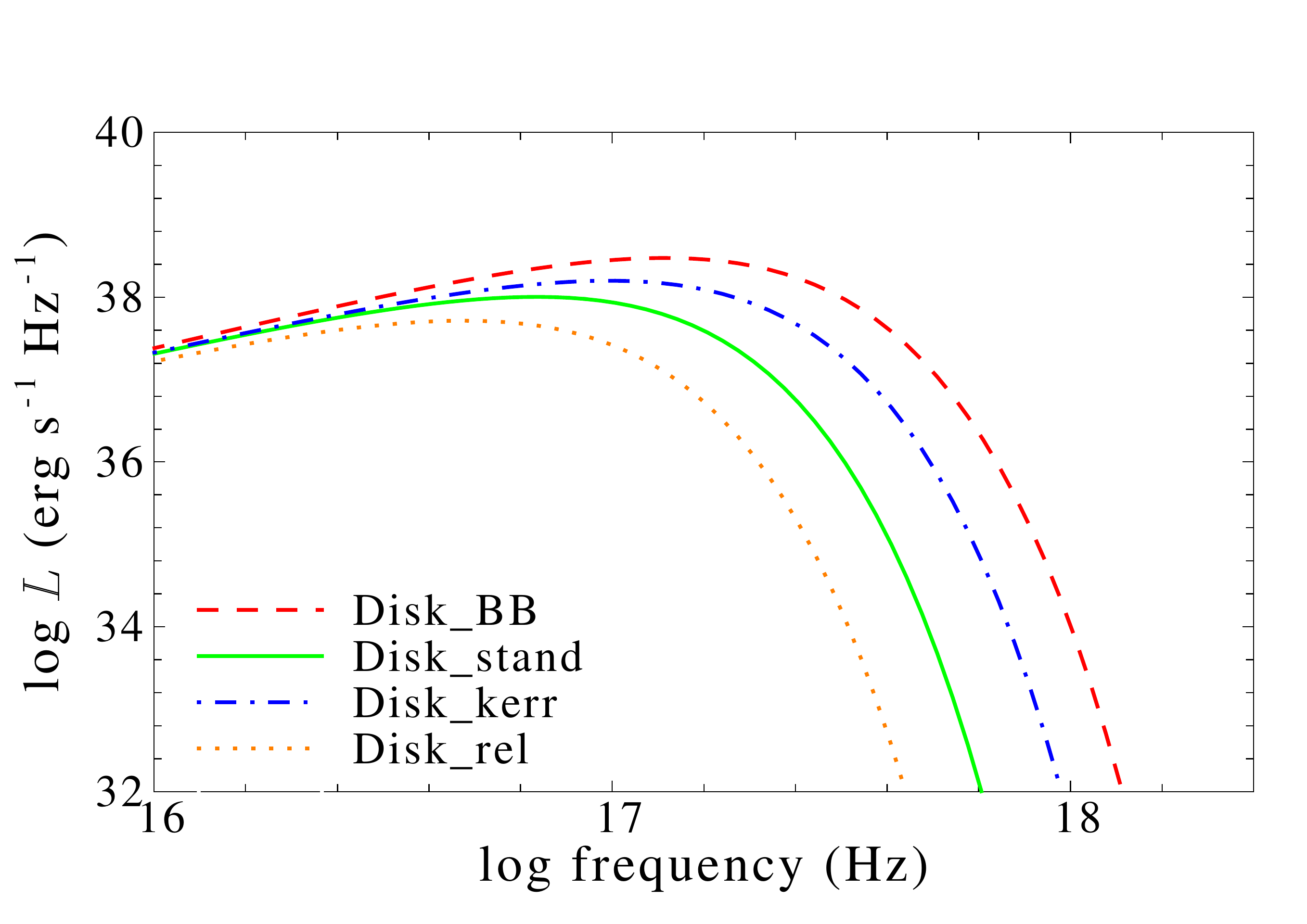}
 \includegraphics[width=84mm]{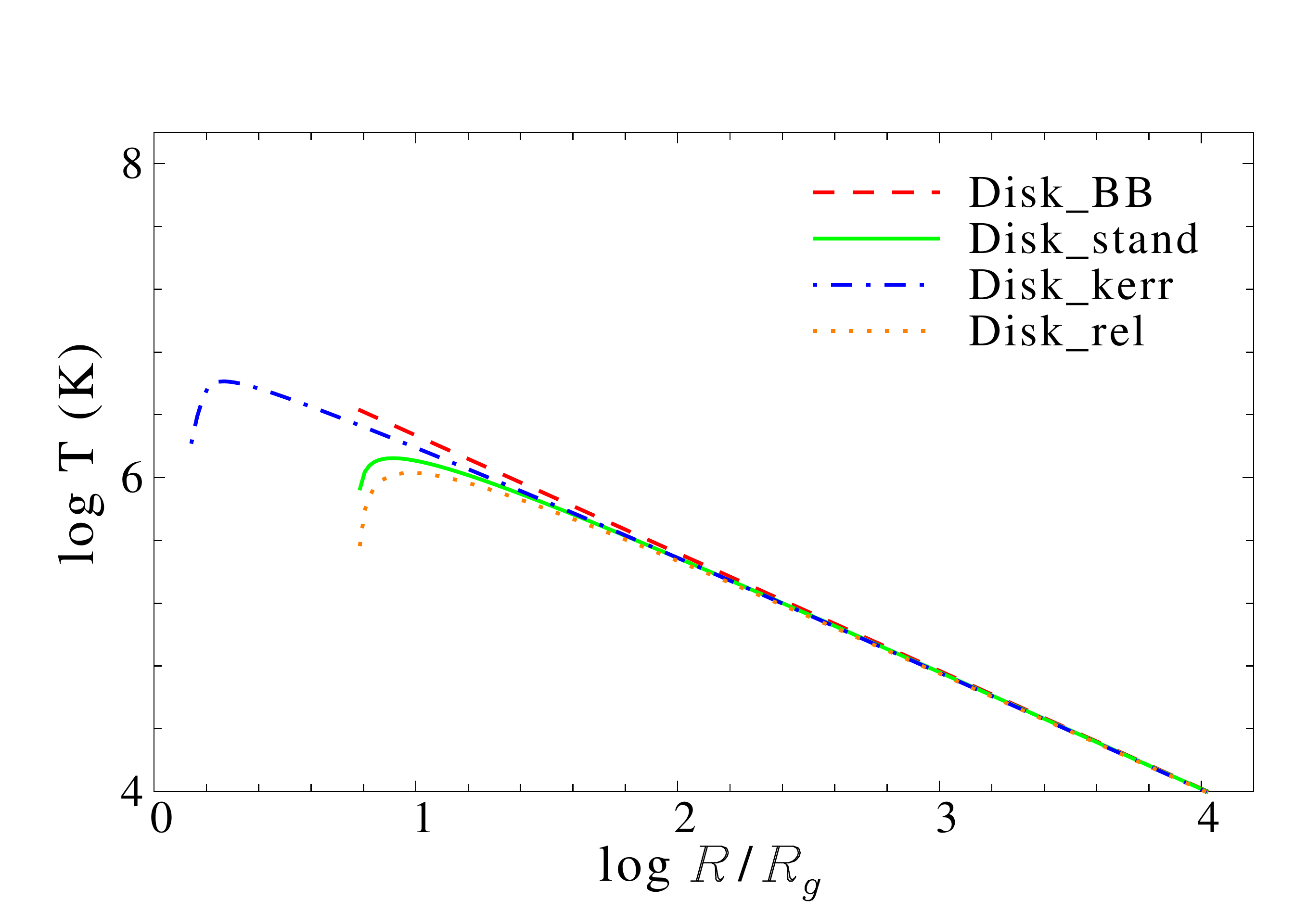}
 \caption{Emitted spectra ({\it top panel}) and temperature profiles ({\it bottom panel}) of the \emph{Disk-BB}, \emph{Disk-stand}, \emph{Disk-rel} 
 and \emph{Disk-kerr} models, obtained for the same characteristic parameters: $M=100M_{\odot}$, $\dot{M}=0.1\dot{M}_{Edd}$, 
 $R_{in,BB}=R_{in}=R_{in,rel}=6R_{g}$, while $R_{in,kerr}$ has been obtained using equation (\ref{A5}). For all models $\cos i=1$.}
 \label{figure4}
\end{figure}
\subsection{\emph{Disk-kerr} model}

The last comparison concerns the \emph{Disk-kerr}/\emph{Disk-BB}
models. If the inclination angle is close to $i=0$ (face-on disk), the
spectral fits of the \emph{Disk-kerr} with the \emph{Disk-BB} at
constant $M$ or $\dot{M}$ are statistically acceptable and return
different values of the inner radius, $R_{in,BB}=7.9R_{g}$ and
$R_{in,BB}=9.2R_{g}$, respectively (see Table~\ref{tab2}). The values
of $\dot{M}_{BB}$ and $M_{BB}$ are larger than the corresponding
values of the \emph{Disk-rel} model (as high as $\sim$ 20$\%$ for
$\dot{M}$ and $\sim$ 25$\%$ for $M$). Also in this case, there are other
fits with fixed $R_{in,BB}$ in the range 7.9--9.2$R_{g}$ for which the
\emph{Disk-BB} and \emph{Disk-kerr} spectra are in good agreement (see
again Table~\ref{tab2}). In Figure~$\ref{figure3}$ we show the results
of one of these fits obtained for $R_{in,BB}\approx 8.9R_{g}$.

We have also tried fits varying the inclination angle $i$. For a
maximally rotating Kerr black hole, the relativistic effects on the
emitted spectrum are significantly stronger than for a Schwarzschild
black hole. The spectrum of the \emph{Disk-kerr} model becomes
intrinsecally different and much harder than the \emph{Disk-BB} one as
$i$ increases, so that it is no longer possible to obtain a
satisfactory fit ($\chi^{2}>2$). For illustrative purposes alone, in
Table~$\ref{tab2}$ we report the values of the fits for $\cos i=0.5$.
\begin{table*}
 \begin{center}
 \begin{minipage}{140mm}
 \caption{Values of the characteristic parameters of the \emph{Disk-kerr} 
 model and those derived by the spectral fits with the \emph{Disk-BB} model 
 for $\cos i=1$ and $\cos i=0.5$. The inner radius of the \emph{Disk-kerr} 
 model has been calculated using equation ($\ref{A5}$) (for $a=0.9981$, 
 $R_{in,kerr}\cong1.36R_{g}$). 
 In the first row $M_{BB}$ is set equal to 
 $M_{Kerr}$; in the second row $R_{in,BB}$ is fixed at an intermediate value
 between 7.9 and 9.2 $R_g$; in the last row 
 ${\dot M}_{BB}$ is set equal to ${\dot M}_{Kerr}$ (for each value of 
 $\cos i$). The results of the spectral fit at fixed radius are shown
 in Figure~\ref{figure3}.
 }
 \label{tab2}
 \begin{center}
 \begin{tabular}{@{}lcccccc}
  \hline
   &  $M_{Kerr}$ & ${\dot M}_{Kerr}$ & $M_{BB}$ & ${\dot M}_{BB}$ & $R_{in,BB}$ & $\chi^{2}_{red}$ \\
   & ($M_{\odot}$) & ($M_{\odot}$) & ($\dot{M}_{Edd}$) &  ($\dot{M}_{Edd}$) & ($R_{g}$) & \\
  \hline
  $\cos i=1$ & 100.0 & 0.1 & 100.0 & 0.075$\pm$0.015 & 7.9$\pm$0.9 & 0.2 \\
   
   & 100.0 & 0.1 & 88.0$\pm$18.0 & 0.095$\pm$0.02 & 8.9 & 0.2 \\
   
   & 100.0 & 0.1 & 85.0$\pm$20.0 & 0.1 & 9.2$\pm$1.3 & 0.2 \\
  \hline
  \hline
  $\cos i=0.5$ & 100.0 & 0.1 & 100.0 & 0.04$\pm$0.015 & 2.4$\pm$1.7 & 2.3 \\
   
   & 100.0 & 0.1 & 60.0$\pm$5.0 & 0.1 & 4.0$\pm$0.4 & 2.2 \\
  \hline
  \end{tabular}
  \end{center}
  \medskip

  \end{minipage}
  \end{center}
\end{table*}

\subsection{Comparison with fitting results from XSPEC}

In order to test our approach, we simulated disk spectra for both 
a Schwarzschild BH and a maximally rotating Kerr BH using the \emph{kerrBB}
model in XSPEC (v. 12.3.0). This model implements the relativistic structure 
equations and the relativistic effects on the emission and propagation of 
radiation (\citealt{b38}). We fitted the \emph{kerrBB} spectra with a 
\emph{diskBB} model (the XSPEC implementation of the MCD) and used 
the parameters of the fit to estimate the mass and accretion rate using 
equations~(\ref{A3}) and~(\ref{A4}). 

Spectra were simulated using the response matrices of the EPIC-pn camera 
onboard {\it XMM-Newton}.
%Standard choices were made for some of the 
%\emph{kerrBB} parameters (Keplerian disk with zero torque at the 
%inner boundary; self-irradiation and limb-darkening switched off).
For the sake of comparison, in the \emph{kerrBB} model we adopted a zero 
torque inner boundary condition and switched off the effects
of self-irradiation and limb-darkening. The disk was assumed to be face-on 
($i=0$) and the color correction factor $f$ was set equal to 1. A fixed distance
of 5 Mpc was assumed. For the mass of the BH and the accretion rate we
chose $100 M_\odot$ and $0.485 \dot M_{Edd}$, respectively, while 
the BH angular momentum $a$ was set equal to 0 (Schwarzschild) or 0.9981 (Kerr).

For a  \emph{kerrBB} spectrum with $a=0$, the best fitting MCD parameters 
are: $T_{in}=0.122\pm 0.002$ keV and $K=39\pm 6$ ($2\sigma$ errors).
From these parameters, assuming $b=19.2\pm 3.5$ (see Table~\ref{tab1}), 
we obtain $M_{BB}=(110\pm 29) M_{\odot}$ (eq.~[\ref{A3}]), in good agreement with 
the assumed value of the BH mass. Similarly, assuming $b=25\pm 6.7$, 
it is $\dot M_{BB}= (0.4\pm 0.28) \dot M_{Edd}$ (eq.~[\ref{A4}]), which is also 
in agreement with the assumed value for the accretion rate.
For a \emph{kerrBB} spectrum with $a=0.9981$, the best fitting MCD parameters
are: $T_{in}=0.248\pm 0.003$ keV and $K=6\pm 0.4$  ($2\sigma$ errors). 
From equations~(\ref{A3}) and~(\ref{A4})
and the values of $b$ reported in Table~\ref{tab2}, we obtain 
$M_{BB}=(105\pm 15) M_{\odot}$ and $\dot M_{BB}=(0.4\pm 0.14) \dot M_{Edd}$,
again in satisfactory agreement with the parameters assumed in simulating
the spectrum.
\begin{table*}
 \begin{center} \begin{minipage}{140mm} \caption{Values of the BH mass $M$ and accretion rate $\dot{M}$ for NGC 1313 X-1 and NGC 1313 X-2,
 obtained using the parameters of the MCD that best-fits their soft components ($K_{BB}=28_{-5}^{+5}$, $T_{in}=0.22$ keV
 for NGC 1313 X-1 ; $K_{BB}=6.66_{-3.5}^{+16}$, $T_{in}=0.25$ keV for NGC 1313 X-2) and different values of $b$ that correspond
 to the different accretion disk models (see Tables~\ref{tab1},~\ref{tab2} and
 section 4; \citealt{b44} and \citealt{b44a} adopted $b=9.5$). The assumed
 distance of NGC 1313 is $3.7$ Mpc (\citealt{b61a}) and the color correction
 factor is $f=1.7$ (\citealt{b59,b73a,b36d}).}  
 \label{tab3}
 \begin{center} 
 \begin{tabular}{@{}lcccccc} 
 \hline 
  & & NGC 1313 X-1 & & NGC 1313 X-2 \\ 
 \hline 
  & $R_{in,BB}$ ($R_{g})$ & $M$ ($M_{\odot}$)
  & $\dot{M}$ ($\dot{M}_{Edd}$)& $M$ ($M_{\odot}$) & $\dot{M}$
   ($\dot{M}_{Edd}$) \\ 
 \hline 
   & 6 & 630$_{-60}^{+50}$ & 0.5$_{-0.1}^{+0.01}$ & 310$_{-100}^{+260}$ & 0.4$_{-0.1}^{+0.3}$ \\ 
   \hline 
   & 7.9 & 480$_{-40}^{+40}$ & -- & 240$_{-80}^{+190}$ & -- \\
   
   & 9.2 & -- & 1.1$_{-0.1}^{+0.1}$ & -- & 0.9$_{-0.2}^{+0.8}$ \\
 \hline 
   & 9.5 & 400$_{-40}^{+40}$ & 1.2$_{-0.1}^{+0.1}$ & 200$_{-70}^{+160}$ & 1.0$_{-0.3}^{+0.8}$ \\ 
 \hline 
   & 13 & 290$_{-30}^{+30}$ & -- & 145$_{-50}^{+120}$ & -- \\

   & 14.9 & -- & 3.0$_{-0.4}^{+0.2}$ & -- & 2.4$_{-0.8}^{+2.0}$ \\
  \hline
   & 19.2 & 200$_{-20}^{+20}$ & -- & 100$_{-30}^{+80}$ & -- \\
   
   & 25 & -- & 8.3$_{-0.8}^{+0.6}$ & -- & 6.7$_{-2.0}^{+5.7}$ \\
  \hline
  \end{tabular}
  \end{center}
  \medskip
  
  \end{minipage}
  \end{center}
\end{table*}
\begin{table*}
 \begin{center}
 \begin{minipage}{140mm}
 \caption{Values of the BH mass $M$ and accretion rate $\dot{M}$ for M81 X-9 and NGC 4559 X-7,
 obtained using the parameters of the MCD that best-fits their soft components ($K_{BB}=20_{-10}^{+20}$, $T_{in}=0.26$ keV for M81 X-9 (obs. 1); 
 $K_{BB}=60_{-40}^{+70}$, $T_{in}=0.21$ keV for M81 X-9 (obs. 2); $K_{BB}=158_{-107}^{+340}$, $T_{in}=0.148$ keV for NGC 4559 X-7)
 and different values of $b$ that correspond to the different accretion disk
 models (see Tables~\ref{tab1},~\ref{tab2} and section 4; \citealt{b44} and
 \citealt{b44a} adopted $b=9.5$). The assumed distance is 9.69 Mpc for NGC 4559
 X-7 (\citealt{b56c}) and 3.4 Mpc for M81 X-9 (\citealt{b24a,b26b}). The color
 correction factor is $f=1.7$ (\citealt{b59,b73a,b36d}).}
 \label{tab4}
 \begin{center}
 \begin{tabular}{@{}lcccccc}
  \hline
    & M 81 X-9 (obs. 1) & & M 81 X-9 (obs. 2) & & NGC 4559 X-7 \\
  \hline
    $R_{in,BB}$ ($R_{g})$  & $M$ ($M_{\odot}$) & $\dot{M}$ ($\dot{M}_{Edd}$)& $M$ ($M_{\odot}$) & $\dot{M}$ ($\dot{M}_{Edd}$) & $M$ ($M_{\odot}$) & $\dot{M}$ ($\dot{M}_{Edd}$)\\
  \hline     
    6 & 490$_{-140}^{+210}$ & 0.7$_{-0.2}^{+0.3}$ & 860$_{-370}^{+400}$ & 0.6$_{-0.2}^{+0.3}$ & 4000$_{-1750}^{+3000}$ & 0.6$_{-0.3}^{+0.5}$\\
    \hline 
      7.9 & 380$_{-110}^{+150}$ & -- & 650$_{-280}^{+300}$ & -- & 3000$_{-1300}^{+2300}$ & -- \\
    
      9.2 & -- & 1.7$_{-0.5}^{+0.7}$ & -- & 1.3$_{-0.6}^{+0.5}$ & -- & 1.4$_{-0.6}^{+1.1}$\\
  \hline     
     9.5 & 310$_{-90}^{+130}$ & 1.8$_{-0.5}^{+0.8}$ & 540$_{-230}^{+260}$ & 1.4$_{-0.6}^{+0.6}$ & 2500$_{-1100}^{+1950}$ & 1.5$_{-0.6}^{+1.2}$\\
  \hline 
     13 & 230$_{-70}^{+90}$ & -- & 400$_{-170}^{+180}$ & -- & 1800$_{-750}^{+1400}$ & --\\

     14.9 & -- & 4.5$_{-1.3}^{+1.8}$ & -- & 3.3$_{-1.4}^{+1.5}$ & -- & 3.8$_{-1.6}^{+2.9}$\\
  \hline
     19.2 & 160$_{-50}^{+60}$ & -- & 270$_{-120}^{+120}$ & -- & 1200$_{-500}^{+1000}$ & --\\
   
     25 & -- & 12.5$_{-5.6}^{+5.2}$ & -- & 9.3$_{-3.9}^{+4.3}$ & -- & 10.1$_{-4.6}^{+8.2}$\\
  \hline
  \end{tabular}
  \end{center}
  \medskip
  
  \end{minipage}
  \end{center}
\end{table*}
\section{Estimates of $M$ and $\dot{M}$ from the X-ray soft component of ULXs}

A certain number of ULXs show two component X-ray spectra that can be
adequately modelled with a soft thermal component plus a power-law. If
the soft component is fitted with a MCD, the fitting parameters (in
particular the normalization $K_{BB}$) return very large values for
the BH mass, often in excess of several hundreds $M_\odot$
(e.g. \citealt{b44,b44a}). These masses are calculated from
equation~(\ref{A3}) assuming specific values for the inner radius $b$
(in units of $R_g$) and the color correction factor $f$. \cite{b44}
and \cite{b44a} adopted $b=9.5$ and $f=1.7$.

In the previous Section, we performed fits of the standard and
relativistic disk with a MCD keeping the BH mass fixed and determined
the (range of) values of $b$ for which the spectra are in statistical
agreement. These values are reported in Tables \ref{tab0},
\ref{tab1} and \ref{tab2} and differ from what assumed by \cite{b44}
and \cite{b44a}. For the standard disk, there appears to be
more agreement with what reported by \cite{b36} and \cite{m00}.
We are then in the position to revise previous estimates of the
BH mass (and accretion rate) of ULXs based on MCD spectral fits of the
soft component, adopting the appropriate value of $b$ for both a
standard and a relativistic disk.

For illustrative purposes, we use the results of the MCD spectral fits
of NGC 1313 X-1, NGC 1313 X-2, M 81 X-9 and NGC 4559 X-7 obtained by
\citet{b44, b44a}, \cite{b73} and \cite{b10a}. From
equations~($\ref{A3}$) and~($\ref{A4}$) of Section~\ref{Standard
accretion disk and MCD}, we calculate the values of $M$ and $\dot{M}$
for different values of $b$ and assuming $f=1.7$, and report them in
Tables~\ref{tab3} and ~\ref{tab4}.
The values that correspond to $b=9.5$ are those reported by \cite{b44}
and \cite{b44a}, whereas the values for $b=7.9, 13, 19.2$ correspond to
those expected for a standard disk around a maximally rotating Kerr, newtonian, 
or Schwarzschild BH, respectively. As can be seen from the Tables, the values of 
the mass are systematically lower (up to a factor 3) than those of a MCD with 
$R_{in,BB}$ at the ISCO ($b=6$). This is clearly a consequence 
of the larger fitting radius required by the newtonian and relativistic
disks. Comparing the masses of the different ``physical'' models,
it is possible to see that the largest value is obtained for a disk around 
a maximally rotating Kerr BH (\emph{Disk-kerr}; see Tables~\ref{tab3} 
and~\ref{tab4}). This is consistent with the fact that the (Boyer-Lindquist) 
radial coordinate of the ISCO for a maximally rotating, Kerr BH is 
$\simeq 1/6$ that of a Schwarzschild BH of the same mass (cmp. 
\citealt{b36c}). Indeed, for a fixed value of $R_{in,BB}$
(in cm) as returned by the MCD fit of an observed spectrum, a lower
value of $b$ necessarily implies a larger mass as $b=R_{in,BB}/R_g \propto
1/M$. 
%as the radius normalized to $R_g$ must be smaller, 
We need to fill the same physical radius 
$R_{in,BB}$ with a bigger BH in order to reduce the value of $b$. 
Ultimately, the smaller value of $b$ 
returned by the fit of the \emph{Disk-kerr} model is a consequence of the 
fact that the Kerr spectrum is significantly harder than those of
the other ``physical'' disk models computed with the same value of the
parameters (see Figure~\ref{figure4}). 

It is also interesting to note that, apart from NGC 4559 X-7, the BH masses 
for a Schwarzschild disk are no longer significantly in 
excess of $\sim 100 M_\odot$, with a distribution clustering in the
interval 100-200 $M_\odot$ (see again Tables~\ref{tab3} 
and~\ref{tab4}). This is true also for M 81 X-9, for which the second
observation returns somewhat larger values of the mass. However, as the limits inferred from the two 
observations of this source must be consistent within the errors, the mass 
of the BH must be in the interval $150-220 M_\odot$ (smaller/larger values
would not be consistent with the second/first observation;
see the last row in Table~\ref{tab4}).
All the mass estimates obtained for the \emph{Disk-rel} model are then
significantly lower than those reported by \cite{b44} and \cite{b44a}.
The only exception is NGC 4559 X-7 for which, however, there are hints of 
timing features in the power density spectrum and it has been proposed
that it may contain a BH of several hundreds solar masses (e.g. \citealt{b10a}). 
The inferred BH masses can reach the values estimated by \cite{b44a} only 
for a disk around a maximally rotating, Kerr BH.

%We note that the value of the accretion rate for some of the relativistic 
%models (the \emph{Disk-rel} model) is close to the critical accretion
%ate at which $L\approx L_{Edd}$ and should be taken with care, as
%in this regime the structure of the accretion disk changes.
%%We are probably at the limit of applicability of the standard disk.
%However, for the \emph{Disk-kerr} model the inferred values of ${\dot M}$ 
%are near Eddington and our results are fully consistent.
%
%
%
%
%%%%%%%%%%%%%%%%%%%%%%%%%%%%%%%%%%%%%%%%%%%%%%%%%%%%%%%%%%%%%%%%%%%%%%%%%%
%
%
%
%
\section{Conclusions}
\label{Conclusions}

The spectrum of a standard disk can be well reproduced by a MCD
model. However, as already noted by \citet{b36,b36a,b36b}, if one
tries to exploit the MCD model to derive the values of the accretion
rate and black hole mass, with the usual assumption that, in the soft
state, the disk terminates at the ISCO ($R_{in,BB}=6R_g$), the
inferred values of $M$ and ${\dot M}$ are largely overestimated. On
the other hand, the spectrum of the standard disk can be fitted by a
MCD model with the same values of $M$ or ${\dot M}$ if we take
$R_{in,BB}=b R_g$ with $b \simeq 13$ or $b \simeq 15$,
respectively. This is in substantial agreement with the approximate
value ($b=15.6$) that can be derived using analytic arguments.
%Therefore, with these values of $b$, the MCD normalization and
%temperature can be safely used to provide a reliable estimate of the
%mass of black hole and of the accretion rate of a standard disk.

Also in the case of a relativistic disk around a Schwarzschild black
hole, for sufficiently small inclination angles the emitted spectrum
can be well reproduced by a MCD model (see also \citealt{b15} and
\citealt{b36b}). For small angles, in fact, one can try to guess the
disk and black hole parameters using the best-fitting MCD model but,
as for the standard disk, a suitable assumption must be made on the
inner disk radius. We found that there exist precise values of
$R_{in,BB}$ ($\simeq 19.2 R_g$ and $\simeq 25 R_g$) for which the
values of $M$ and ${\dot M}$ inferred from the MCD are the same as
those provided by the relativistic disk. For large angles, owing to
genuine relativistic effects, no satisfactory MCD spectral fit of the
relativitisc disk spectrum can be obtained (e.g. \citealt{b15}).
%and hence, any correct assessment of disk and black hole physical
%parameters must rely on adopting the appropriate relativistic model.

The spectrum of an accretion disk around a fast rotating Kerr black
hole, is in general more strongly affected by relativistic effects
compared with those of a disk around a Schwarzschild blak hole. As a
consequence, it can be well reproduced by a MCD model, only in the
case in which the disk is essentially face-on. In this assumption,
there exist values of $R_{in,BB}$ for which the MCD and Kerr disks return
similar values of $M$ and ${\dot M}$ ($R_{in,BB}\simeq 7.9 R_g$ and
$R_{in,BB}\simeq 9.2 R_g$, respectively). Also in this case, for large
inclination angles, the spectrum of the \emph{Disk-kerr} model becomes
intrinsecally different and much harder than the \emph{Disk-BB} one,
so that it is no longer possible to obtain a satisfactory fit with the
MCD model.

%\section{Discussion}
%\label{Discussion}

%In this paper, we compared the temperature profiles and spectra of a
%standard and relativistic accretion disk with those of a simple MCD
%model. 

The model-model comparison approach adopted in this Paper was prompted 
by the idea to determine the parameter space in which the
MCD is in agreement with ``more physical'' standard accretion disk
models and provide ``recipes'' for adjusting the estimates of the disk
inner radius, the black hole mass and the accretion rate. In fact,
despite the uncertainties related to radiative transfer and to the
physical state of the inner disk, the parameters of the MCD fit are
often used for inferring the mass and accretion rate in X-ray
binaries, in particular in ULXs. The results of the comparison that
we performed can be used to revise estimates of $M$ and ${\dot M}$ 
obtained in this way. This has also been tested by applying
our ``recipe procedure'' to the results from of a \emph{diskBB} 
fit to simulated \emph{kerrBB} spectra performed directly in XSPEC.

We considered the case of a few ULXs (NGC 1313 X-1, NGC 1313 X-2, M81
X-9 and NGC 4559 X-7) for which MCD spectral fits of their X-ray soft
spectral components have been published and/or values of the BH mass
estimated (\citealt{b10a,b44,b44a,b73}). From the parameters of the
fit with the \emph{Disk-BB} ($T_{in}$ and $K_{BB}$) we found that,
assuming that the inner disk boundary of the MCD is at or close to
$6R_{g}$, the values of $M$ and $\dot{M}$ can be severely
overestimated. 
Adopting the appropriate value of the inner radius $b$
(in units of $R_g$) for the \emph{Disk-rel} model,
we obtain that the BH masses are in the range
$\approx 100-200M_{\odot}$ for NGC 1313 X-1, NGC 1313 X-2, M81 X-9, and
$700-2000 M_{\odot}$ for NGC 4559 X-7 (see Tables~\ref{tab3} 
and~\ref{tab4}). The relativistic 
disk models have BH masses systematically lower (up to a factor 3) than 
those of a MCD with $R_{in,BB}$ at the ISCO.

As already mentioned, sophisticated relativistic accretion disk models
(e.g. \citealt{b26a,b38,b36d}) that implement the relativistic structure 
equations and the relativistic effects on the emission and propagation of 
radiation, are available in XSPEC and can clearly be used to perform 
accurate fits of observed spectra, in particular for large inclination
angles when the relativistic effects produce significant differences. 
An analysis of this type has recently been performed by \citet{b72} and 
\citet{b36c}. 
It is interesting to note that the BH masses for a sample 
of disk-dominated ULXs obtained by \citet{b36c} are below $\sim 100 M_\odot$.
Even if their best fits require large values of the BH spin, 
BHs of several hundreds solar masses are not needed, in 
substantial agreement with our findings.
%agree well with the range of values of $M$ that we find in our analysis.

Clearly, it  is  possible  that at  least   some ULXs  for  which
curvature  above $\sim$ 2-3 keV is present in the X-ray spectrum
are in a different accretion regime, in which the disk is in a diverse
physical state    (accreting at  or above  the    Eddington rate;  see
e.g. \citealt{b56a}; \citealt{b56b}). In this case, the assumptions of
a standard accretion  disk  break down and different  spectral  models
should be adopted (e.g. \citealt{b61}; \citealt{m07}). However, in the
assumption that they are in a standard accretion regime (below Eddington), we 
found that, in three cases (NGC 1313 X-1, X-2 and M 81 X-9), the black 
hole masses inferred for a standard disk around a Schwarzschild black 
hole are in the interval $\sim 100-200 M_\odot$.
Only if the black hole is maximally rotating are the masses comparable 
to the much larger values previously derived in the literature.

\section{Acknowledgements}

We are indebted with an anonymous referee for his/her valuable comments.
We thank Andrea Martocchia for his careful reading of the manuscript. 
LZ acknowledges financial support from INAF through grant PRIN-2007-26.

%
%
%
%
%%%%%%%%%%%%%%%%%%%%%%%%%%%%%%%%%%%%%%%%%%%%%%%%%%%%%%%%%%%%%%%%%%%%%%%%%%%%%
%
%
%
%

%
%
%
%
%%%%%%%%%%%%%%%%%%%%%%%%%%%%%%%%%%%%%%%%%%%%%%%%%%%%%%%%%%%%%%%%%%%%%%%%%%%%%
%
%
%
%
\end{document}